\documentclass{amsart}

\RequirePackage{ifthen}

\newboolean{oneResultSequence}
\setboolean{oneResultSequence}{false} 

\newcommand{\numberResults}{
\ifthenelse{\boolean{oneResultSequence}}{
\newtheorem{axiom}{Axiom}
\newtheorem{theorem}{Theorem}
\newtheorem{lemma}[theorem]{Lemma}
\newtheorem{proposition}[theorem]{Proposition}
\newtheorem{corollary}[theorem]{Corollary}
\theoremstyle{definition}
\newtheorem{definition}[theorem]{Definition}
\newtheorem{example}[theorem]{Example}
\newtheorem{xca}[theorem]{Exercise}
\theoremstyle{remark}
\newtheorem{remark}[theorem]{Remark}
\newtheorem{conjecture}[theorem]{Conjecture}
}{
\newtheorem{axiom}{Axiom}
\newtheorem{theorem}{Theorem}
\newtheorem{lemma}{Lemma}
\newtheorem{proposition}{Proposition}
\newtheorem{corollary}{Corollary}
\theoremstyle{definition}
\newtheorem{definition}{Definition}
\newtheorem{example}{Example}

\theoremstyle{remark}

}}

\newcommand{\abbrevEnvir}{
\expandafter\newcommand\expandafter{\csname bi\endcsname}{\begin{itemize}} 
\expandafter\newcommand\expandafter{\csname ei\endcsname}{\end{itemize}}
\expandafter\newcommand\expandafter{\csname be\endcsname}{\begin{enumerate}} 
\expandafter\newcommand\expandafter{\csname ee\endcsname}{\end{enumerate}}
\expandafter\newcommand\expandafter{\csname bc\endcsname}{\begin{center}} 
\expandafter\newcommand\expandafter{\csname ec\endcsname}{\end{center}}
}

\renewcommand{\Re}{\mathbb{R}}

\newcommand{\authorGreen}{\ifthenelse{\isundefined{\authorInPageHeader}}
{\author[\relax]{Edward J. Green}}
{\author[\authorInPageHeader]{Edward J. Green}}}

\authorGreen
\address{Department of Economics, The Pennsylvania State University,
  University Park, PA 16802, USA}
\email{eug2@psu.edu}

\newcommand{\theor}[3][]{\begin{theorem}[#1] #3 \label{thm:#2} \end{theorem}}

\def\lema#1#2{\begin{lemma} #2 \label{lem:#1} \end{lemma}}

\def\display#1#2{\begin{equation} #2 \label{eqn:#1} \end{equation}}
\def\eqn#1{(\ref{eqn:#1})}

\newboolean{RANW}
\setboolean{RANW}{false}
\newlength{\testWidth} \newlength{\badWidth}

\newcommand{\useRsltAndNumberwithin}{
\setboolean{RANW}{true}
\settowidth\badWidth{\ref{useRsltAndNumberwihin_generates_spurious_undefined-reference_messages.}}
\newcommand{\badMatch}{\equal{\the\testWidth}{\the\badWidth}}
}

\newcommand{\rslt}[1]{%
\ifthenelse{\boolean{RANW}}%
{%
{\settowidth{\testWidth}{\ref{lem:#1}}\ifthenelse{\badMatch}%
{%
\settowidth{\testWidth}{\ref{prp:#1}}\ifthenelse{\badMatch}%
{%
\settowidth{\testWidth}{\ref{thm:#1}}\ifthenelse{\badMatch}%
{%
\settowidth{\testWidth}{\ref{cor:#1}}\ifthenelse{\badMatch}{\textbf{result ??}}%
{corollary \ref{cor:#1}}%
}%
{theorem \ref{thm:#1}}%
}%
{proposition \ref{prp:#1}}%
}%
{lemma \ref{lem:#1}}%
}}%
{%
\ifthenelse{\ref{lem:#1} > 0}{lemma \ref{lem:#1}}%
{%
\ifthenelse{\ref{prp:#1} > 0}{proposition \ref{prp:#1}}%
{%
\ifthenelse{\ref{thm:#1} > 0}{theorem \ref{thm:#1}}%
{%
\ifthenelse{\ref{cor:#1} > 0}{corollary \ref{cor:#1}}{\textbf{Result ??}}%
}%
}%
}%
}}

\newcommand{\Rslt}[1]{%
\ifthenelse{\boolean{RANW}}%
{%
{\settowidth{\testWidth}{\ref{lem:#1}}\ifthenelse{\badMatch}%
{%
\settowidth{\testWidth}{\ref{prp:#1}}\ifthenelse{\badMatch}%
{%
\settowidth{\testWidth}{\ref{thm:#1}}\ifthenelse{\badMatch}%
{%
\settowidth{\testWidth}{\ref{cor:#1}}\ifthenelse{\badMatch}{\textbf{Result ??}}%
{Corollary \ref{cor:#1}}%
}%
{Theorem \ref{thm:#1}}%
}%
{Proposition \ref{prp:#1}}%
}%
{Lemma \ref{lem:#1}}%
}}%
{%
\ifthenelse{\ref{lem:#1} > 0}{Lemma \ref{lem:#1}}%
{%
\ifthenelse{\ref{prp:#1} > 0}{Proposition \ref{prp:#1}}%
{%
\ifthenelse{\ref{thm:#1} > 0}{Theorem \ref{thm:#1}}%
{%
\ifthenelse{\ref{cor:#1} > 0}{Corollary \ref{cor:#1}}{\textbf{Result ??}}%
}%
}%
}%
}}

\def\Section#1#2{\section{#2\label{sec:#1}}}

\def\sec#1{{s}ection \ref{sec:#1}} 

\newenvironment{enum} 
{\begin{list}{\makebox[\labelwidth][l]{(\arabic{enumi})}}{\usecounter{enumi}}
\setcounter{enumi}{\value{equation}}}
{\setcounter{equation}{\value{enumi}} \end{list}}

\newcommand{\meti}[2]{\item #2 \label{eqn:#1}} 

\newcommand{\eqnref}[1]{\ref{eqn:#1}}

\newcommand{\secref}[1]{\ref{sec:#1}} 

\numberResults

\title[\relax]{J.~S.~Mill's Liberal Principle and Unanimity}

\date{2019 \quad (first draft, 1994)}

\begin{document}

\begin{abstract}
The broad concept of an individual's welfare is actually a
cluster of related specific concepts that bear a ``family resemblance'' to
one another.  One might care about how a policy will affect people both in
terms of their subjective preferences and also in terms of some notion of
their objective interests. This paper provides a framework for evaluation
of policies in terms of welfare criteria that combine these two
considerations. Sufficient conditions are provided for such a criterion to
imply the same ranking of social states as does Pareto's unanimity
criterion. Sufficiency is proved via study of a community of agents with
interdependent ordinal preferences.
\end{abstract}

\maketitle

\Section a{Introduction}

The broad concept of a person's welfare is actually a cluster of
related, specific concepts.  Economists have tended to focus attention
almost exclusively on just one of these specific concepts, which
interprets questions regarding people's welfare as being about their
subjective preferences. One might care also about some notion of
people's objective interests.  I use the term `objective interest' to
refer to any specific notion of a person's well-being that might be
quantified by an index of measurements of various aspects of the
person's situation.

Obviously there can be many distinct notions of a person's objective
interest. For instance, some notions are based exclusively on the
adequacy of a person's access to basic physical necessities, while
other notions encompass social and psychological aspects of the
person's situation as well. (Notably, Rawls's (1971) characterization
of a person's objective interest in terms of ``primary goods''
includes these broader aspects.) It should be possible to design a
satisfactory framework for evaluation of policies in terms of welfare
criteria based on notions of people's objective interests, and also in
terms of criteria that combine considerations of objective interests
and preferences, just as neoclassical economists have succeeded in
doing for preference-based notions.

Moreover, once it is granted that there may several alternative
concepts of welfare that are all worthy of consideration in policy
making, then some interesting general questions arise concerning the
relationships between them. Are there salient features common to
various specific notions of a person's objective interest---analogous
to the ordinal features by which economists characterize preferences
in an abstract, general way---that might serve as a basis for economic
analysis? If so, can any general comparison be made between the
conclusions of welfare analyses based on various notions of objective
interest possessing these features, or between the conclusions of such
an analysis and a neoclassical analysis based exclusively on
preferences?  In particular, are there any interesting conditions
under which all of these various analyses would reach identical
conclusions? This paper studies these questions.

The main goal of the paper is to compare the neoclassical welfare criterion
of unanimous preference (formulated by Pareto (1909) and emphasized in
welfare analysis by Wicksell (1935)) to an alternative criterion that is
somewhat in the same spirit, but that relies in an essential way on a notion
of each person's objective interest as well as on a notion of persons'
preferences. The formulation of this alternative criterion is suggested by
John Stuart Mill's discussion of liberalism.\footnote{Mill's thinking has also
been influential in stimulating the formal investigation (beginning with Sen
1970) of the problem of constraining Paretian welfare analysis to respect
persons' rights. Schick (1980) and Riley (1988) have employed formal
social-choice theory to interpret Mill's political philosophy as a whole.}
Mill emphasizes two connected ideas: that there is a notion of each person's
interest that is distinct from that person's preferences, and that a person
ought to be permitted to do as he pleases so long as he does not damage the
interests (in this non-preferential sense) of others (Mill 1859, Book I and
Book IV). He extends this idea of liberal permissibility to coalitions. That
is, he proposes that a group of people ought to be permitted to act
unanimously as long as they do not damage the interest of anyone outside the
group. This proposal would license actions of which the entire community
approves, because the group that includes everyone would act unanimously and
there would be no outsider whose interest could be damaged. Thus Mill's
criterion of liberally permissible action is at least as broad as the
Pareto-improvement criterion. Furthermore Mill states explicitly that he
interprets the notion of people's interests narrowly enough so that some
actions would also be permissible although they were not unanimously
approved. That is, Mill takes the position that a person's mere preference
that an action should not be taken does not make the action damaging to the
person's objective interest. He proposes that someone should be permitted to
take an action of which no one else approves, as long as those others'
disapproval reflects such mere preferences unconnected to their interests.
On this interpretation, Mill's proposed criterion of liberal permissibility
is strictly broader than the Pareto-improvement criterion.

The preceding discussion makes it clear that Mill's views on liberally
permissible action are based on two separate considerations: a
specific notion of what objective interests people possess, and a
general characterization of the relationship between people's
objective interests and the permissibility of actions that other
people might wish to take.  Thus someone might endorse Mill's general
characterization, but differ with him about concrete cases on the
basis of holding a different view of what are people's objective
interests. (Indeed, someone might suppose contrary to Mill that a mere
preference does create a corresponding objective interest.  When
Mill's general characterization is interpreted according to that view,
the resulting criterion is simply a reformulation of the
Pareto-improvement criterion.) The formal theory to be presented here
will emphasize the separation between Mill's two considerations. The
main result to be proved is that, even in the context of an assumption
about the relationship between preferences and interests that is
significantly weaker than to posit an identification between them,
Mill's general characterization of liberal permissibility coincides
exactly with the Pareto-improvement criterion. This result probably
would have surprised both Mill and his philosophical critics. It shows
that the conflicts between some alternative welfare criteria are much
less pervasive than might have been believed.  However it does
identify some particular situations (namely, those in which the
relationship between preferences and welfare assumed in the theorem
are implausible) in which the liberal and Paretian criteria are likely
to lead to different conclusions.

\Section b{The setting of the problem}

In order to examine carefully the logic behind Mill's suggestion that his
liberal principle permits a wider class of actions than just those that
accomplish unanimously preferred changes, I am going to restate the
suggestion in language that is parallel to that of neoclassical welfare
economics. There a person's preferences are represented formally by a binary
relation that may hold between {\it social states}, which in turn represent
possible situations of the community. A person's objective interest can also
be also be represented formally by a binary relation between social states.
As in the case of preferences, a distinct relation is identified with each
person or {\it agent} in the community.

In welfare economics, the relation of unanimous preference (often called the
``Pareto-improve\-ment relation'') is defined from the preference relations of
individuals. Specifically, one social state is the {\it unanimous successor}
of another, status-quo social state if every agent prefers the change to it
from the status quo, with at least one agent's preference being strict.
Analogously, define a social state to be the {\it liberal successor} of a
status-quo social state if all members of some group of agents prefer the
change to it from the status quo, with at least one member of the group
having a strict preference, and if also the change does not reduce the
welfare of any agent outside the group. (In both of these definitions, I
refer to one social state as the status quo only to distinguish it from the
other social state in the pair. I do not mean to imply that it is
historically determined or special in any other way.)

Using these definitions, Mill's suggestion can be restated as the
thesis that one social state may be a liberal successor of another
without necessarily being its unanimous successor. Now I specify a
formal theory in which this thesis can be expressed. The aim of this
theory is to characterize the logic of the thesis in a way that is
completely explicit, and that is also general in the sense of being
independent of specific proposals regarding what might constitute a
person's interest.

Consider a finite set $I$ of agents, and a set $X$ of social states.
$(i,\,j,$ and $k$ will range over $I,$ and $w,\,x,\,y,$ and $z$ will range
over $X.)$ To each agent $j$ are associated two binary relations on
$X,\;W_{j}$ and $R_{j}.$ When $xW_{j}y,$ this means that state $x$ provides
for the objective interest of agent $j$ as well as state $y$ does. I will
refer to $W_j$ as the {\it interest relation,} or simply the {\it interest,}
of agent $j.$ When $xR_{j}y,$ this means that $j$ weakly prefers $x$ to $y.$

Define connectedness of a relation to mean that each pair of states is
related in at least one direction, and for each agent $j,$ assume that

\begin{enum}
\meti a{$W_{j}$ is transitive and reflexive, and $R_{j}$ is transitive and
connected.}
\end{enum}

\noindent $V_{j}$ and $P_{j}$ will denote the asymmetric (or strict) parts
of $W_{j}$ and $R_{j}$ respectively. $E_{j}$ and $I_{j}$ will denote the
symmetric parts of $W_{j}$ and $R_{j}$ respectively.

Within this simple formal language, it is possible to state three
substantive conditions that may be placed on agents' preferences. The first
condition actually guarantees three things: that each agent cares only about
his own interest and possibly the interests of others, that each agent
prefers a social state that better provides for his own interest  if the
interests of others are held constant, and that each interest relation is
connected on any set of states on which all other agents' interests are
constant. Formally, for any agent $j$ and for any states $x$ and  $y,$

\begin{enum}
\meti b{If $\forall i\neq j\;xE_{i}y,$ then $[xW_{j}y \iff xR_{j}y].$}
\end{enum}

\noindent This will be referred to as the condition that {\it preferences
are based on interests.}

Second, agent $j$ will be called {\it nonpaternalistic} if his
preferences are consonant with his own interest and the preferences of
others. Formally, $j$ is nonpaternalistic if, for all $x$ and
$y,$\footnote{The use of `nonpaternalism' for this condition is well
  entrenched in the literature on the interdependent preferences. It
  is easy to think of examples in which the condition applies in ways
  that intuitively have nothing to do with nonpaternalism, though. For
  example, Sen (1970), considers a prude who would prefer to be the
  one to read {\sl Lady Chatterley's Lover} if anyone has to read it
  at all,and someone else who would like to read the book but who
  maliciously would even more enjoy forcing the prude to read
  it. Regarding someone who prefers that his own welfare should be
  maximized \emph{certeris paribus,} the condition formulated here is
  necessary, but not sufficient, for the person's preferences to be
  genuinely non-paternalistic.}

\begin{enum}
\meti c{If $xW_{j}y$ and $\forall i\neq j\;xR_{i}y,$ then $xR_{j}y.$}
\end{enum}

The third condition that can be stated regarding an agent's preferences is
that his preferences about trade-offs between the interests of agents within
any one group in the population are independent of the situation regarding
interests of other agents outside the group. Formally, agent $i$ has
preferences that are {\it separable in interests} if for any partition of
$I$ into groups $J$ and $K$ and for any states $w,\,x,\,y,$ and $z,$

\begin{enum}
\meti d{If $\forall j\in J\;[wE_{j}y$ and $xE_{j}z]$ and $\forall k\in
K\;[wE_{k}x$ and $yE_{k}z],$\\ then $wR_{i}x \iff yR_{i}z.$}
\end{enum}

Separability of preferences in interests will play an important
technical role in the arguments below. It is a necessary condition for
an agent's preferences to be representable by a utility function that
is a sum of functions that in turn are closely related to his and
other agents' interest relations. Given the other assumptions that
have been made here, and a few others that will be introduced shortly,
it is also sufficient for the existence of such a representation. It
should be noticed, though, that separability is an assumption with a
lot of substantive content. For example, if each agent's interest is
taken to be his wealth, then it rules out the possibility that an
agent prefers small increases in wealth for others whenever they are
poorer than he is, but that (perhaps because of envy) he prefers that
they should suffer small decreases in wealth when they are richer than
he is. This paper will conclude with an example that shows that the
liberal and unanimity relations may coincide even when agents'
preferences are not separable in interests.

Three technical assumptions about the topological structure of the set of
social states, and of agents' interest and preference relations, will also
be needed. In contrast to the separability assumption, these assumptions do
not seem to raise significant issues of interpretation. The meanings of the
first two assumptions are immediately obvious. The third assumption states
that if $E_{j}$ is viewed as a correspondence, then it is lower
hemicontinuous:

\begin{enum}
\meti e{$X$ is a connected, separable topological space.}

\meti f{For every agent $j,\;W_{j}$ and $R_{j}$ have closed graph in
$X^{2}.$}

\meti g{For every agent $j,$ pair of states  $y$ and $z,$ and neighborhood
$Z$ of $z,$ if $yE_{j}z,$ then there exists a neighborhood $Y$ of $y$ such
that $\forall y'\in Y\;\exists z'\in Z\; y'E_{j}z'.$}
\end{enum}

The unanimity and liberal relations now have to be defined. State $x$ is a
{\it unanimous successor} to $y$ (and is also described as being {\it Pareto
superior} to $y$ or a {\it Pareto improvement} of $y$) if

\display h{\forall i\; xR_{i}y \text{ and } \exists j\;xP_{j}y.}

Define $x$ to be a {\it liberal successor} of $y$ if, for some coalition
$J\subseteq I,${}\footnote{This relation of liberal succession is one of two
possible ways (within the present theory) to formalize Mill's criterion for
when a change of social state should be countenanced. The other possibility
would be to stipulate that a coalition may make any change of state that
does not give non-members strictly lower levels of welfare than they
previously enjoyed. That is, $x$ could be defined to be a liberal successor
of $y$ if, for some coalition $J\subseteq I,$
\begin{itemize}
\item[(\eqnref{i}$'$)] $\forall i\in J\; xR_{i}y$ and $\exists j\in J\; xP_{j}y$ and not $\exists
k\not\in J\; yV_{k}x.$
\end{itemize}
In general, $(9')$ defines a more
permissive (i.e., set-theoretically larger) relation than does \eqn i.
However, \eqn i and $(9')$ define the same relation if all welfare relations
$W_{j}$ are connected. While connectedness is not assumed directly, it is
implied by other assumptions that will be used here (cf. lemma 6 below).}

\display i{\forall i\in J\; xR_{i}y \text{ and  } \exists j\in J\;
xP_{j}y \text{ and } \forall k\not\in J\; xW_{k}y.}

\Section c{Mill and Pareto contrasted}

At the close of \sec a, I provided an informal summary of Mill's liberal
principle. I pointed out the principle always countenances Pareto
improvements. The formal theory just set forth affords an explicit proof of
this. In particular, taking $J=I$ in \eqn i yields \eqn h, showing that the
unanimous-succession relation is a subrelation of the liberal-succession
relation. This is exactly the formal statement of the assertion.

I also mentioned Mill's thesis that his principle countenances some
changes of social state that are not Pareto improvements. I emphasized
that the thesis follows from his view that there is at most a very
loose relationship between a person's preferences and his objective
interest or even people's objective interests in general. However, in
\sec b I have assumed that all agents have nonpaternalistic
preferences that are based on interests and separable in
interests. These conditions all relate preferences to interests. One
might wonder whether these relationships restrict the scope of the
liberal principle so tightly that it must coincide exactly with the
Pareto principle. Because the conditions in \sec b have been specified
in an explicit and formal way, it is possible to construct an example
that shows definitely that the conditions {\it do not imply} that
Mill's principle and the Pareto principle coincide. I will provide one
such example next. This example should not  necessarily be interpreted
to show that Mill's thesis is incorrect, though. Rather it shows that the
conditions introduced in \sec b are weaker in some circumstances than
they might seem to be. After presenting the example. I will introduce
some further assumptions that rule out such degenerate
circumstances. Once these assumptions have been set forth, the stage
will be set to examine the relationship between Mill's liberal
principle and the Pareto principle. Now, here is the example of a
community in which the liberal-succession relation is much more
inclusive than the unanimous-succession relation.

Let $I=\{1,\,2,\,3,\,4\},$ and let $X$ be an interval of the real line.
Suppose that $xR_{1}y$ and $xW_{2}y$ if $x\geq y,$ and that $xW_{1}y$ and
$xR_{2}y$ if $ x \le y.$ Let agent 3 be identical to agent 1, and agent 4 to
agent 2. It is easily verified that all of the assumptions \eqn a--\eqn g
are satisfied. However, no pair of social states is related by the
unanimous-succession relation, but every pair of distinct social states is
related by the liberal-succession relation.

In this example, the assumptions \eqn b and \eqn c of interest-based
preference and nonpaternalism have been trivialized. That is, each
assumption is an implication with a hypothesis that is never satisfied. Now
I will define four conditions which are intended to identify environments
that are rich enough for the two assumptions to have nontrivial content.
Note that the example above does not satisfy any of these conditions.

First, agents' interests will be said to have {\it product structure}
if$\;$\footnote{It would be desirable to make only an assumption to the effect
that there is some latitude for agents' interests to vary independently. The
present, inordinately strong, assumption is made in order to invoke a
theorem of conjoint measurement theory below. Krantz, et. al. ((1971), pp.
275-6) discuss the general need to have a theory of measurement based on a
weaker assumption. Thus the objectionable feature of the present assumption
seems to be a convenient technical fiction that need not be viewed as
limiting the validity of the results to be derived here. Of course, social
states must be viewed as conceptually possible profiles of interests that
are not necessarily technically feasible to satisfy all at once.}

\display j{\forall \{ x_{i}\} _{i\,\in\,I}\;\exists y\;\forall i\;
yE_{i}x_{i}.}

Second, agent $i$ will be said to have {\it idiosyncratic interest\/} if 

\display k{\exists x\; \exists y\; [xV_{i}y \text{ and  } \forall j\neq i\;
xE_{j}y].}

Third, analogously, agent $i$ will be said to have {\it idiosyncratic
preferences\/} if

\display l{\exists x\; \exists y\; [xP_{i}y \text{ and } \forall j\neq i\; xI_{j}y].}

Fourth, it will be said that an {\it unambiguous improvement\/} is possible
if

\display m{\exists x\; \exists y\; \forall i\; [xV_{i}y \text{ and } xP_{i}y].}

\Section d{Agents' judgments and interpersonal\\ agreement regarding
interests}

Taken together, assumptions \eqn b that agents' preferences are based on
interests and \eqn d that agents' preferences are separable in interests
have an interpetation that each agent's preferences are based on
considerations of trade-offs between his own interest and other agents'
interests. (There might well be other possible interpretations of agents'
preferences as well.) The idea of making trade-offs among interests implies
that the difference between the degree to which an agent's interest is
satisfied in two distinct states is conceived as being a cardinal magnitude.
Moreover, agents may either agree or disagree with one another about
comparisons between these magnitudes. This is a subtle point, and it is also
an important one for the welfare analysis which is to follow. Therefore let
us consider it carefully now.

Consider two agents $i$ and $j$, each of whose interest is completely
determined by what kind of house he lives in. Suppose that there are four
types of house---hovels, cottages, mansions, and palaces---in increasing
order of satisfaction of an agent's interest. Suppose that a social state is
simply a specification of a type of house for each agent.

The ordinal ranking of types of house determines the answers to some
questions regarding cardinal comparisons, such as ``Would it make more
difference for the satisfaction of agent $j\/$'s interest if he were to move
from a cottage to a palace than if he were to move from a cottage to a
mansion?'' Clearly the former difference is larger, since the two changes
begin at the same point but the latter change accomplishes only part (in
terms of the ranking) of what the former accomplishes. However, the ranking
does not determine the answers to other questions such as ``Would it make a
bigger difference for the satisfaction of agent $j\/$'s interest if he were
to move from a mansion to a palace, than if he were to move from a hovel to
a cottage?'' That is because neither of the ordinal intervals described by
these two changes includes the other. If one were to represent the
magnitudes of differences in satisfaction of an agent's interest as
distances between points on a line, then pictures like either of the
following would be possible.

\bigskip\bigskip\bigskip { \def\h{\raise .5ex\hbox{$[\rm hovel ]$}}
\def\c{\raise .5ex\hbox{$[\rm cottage ]$}} \def\m{\raise .5ex\hbox{$[\rm
mansion ]$}} \def\p{\raise .5ex\hbox{$[\rm palace ]$}}

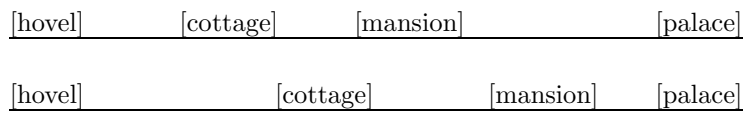
$$ \vtop{\hbox{\h \hskip .5in \c \hskip .4in \m \hskip 1in \p}\hrule
\bigskip\bigskip \hbox{\h \hskip 1in \c \hskip .6in \m \hskip .3in
\p}\hrule} $$

\bigskip\bigskip\bigskip }

Information about agents' preferences can be interpreted to represent the
agents as implicitly answering such questions, though. For example, consider
the following two pairs of states: $x$ and $y$, and $x^\prime$ and
$y^\prime$.

\bigskip

{ \def\state#1#2#3{State $#1$: & Agent $i\/$ lives in a #2; & Agent $j\/$
lives in a #3.\cr}

\halign{\indent #\hfil &\qquad #\hfil &\quad  #\hfil\cr

\state{x\phantom{{}^\prime}}{mansion}{mansion}
\state{y\phantom{{}^\prime}}{cottage}{palace} }

\medskip

\halign{\indent #\hfil &\qquad #\hfil &\quad  #\hfil\cr

\state{x^\prime}{mansion}{hovel} \state{y^\prime}{cottage}{cottage} } }

\bigskip

Suppose that agent $i\/$ strictly prefers state $y\/$ to state $x\/$
and also strictly prefers state $x^\prime\/$ to state
$y^\prime\/$. Then I will interpret agent $i\/$ as judging that agent
$j\/$'s interest is affected more significantly by a move from a
mansion to a palace than by a move from a hovel to a cottage, in the
following sense. Agent $i\/$ would be willing to bear the sacrifice of
having to move from a mansion to a cottage in order to enable $j\/$ to
move from a mansion to a palace, but he would be unwilling to bear the
same sacrifice in order to enable $j\/$ to move from a hovel to a
cottage. This pair of preferences can be interpreted as reflecting a
judgment on the part of $i\/$ that a move from a mansion to a palace
would make a larger difference for the satisfaction of $j\/$'s
interest than would a move from a hovel to a cottage.

In an environment with many social states, this judgments-about-interests
interpretation of agents' preferences will have restrictive implications.
For example, consider another two pairs of social states.

\bigskip

{ \def\state#1#2#3{State $#1$: & Agent $i\/$ lives in a #2; & Agent $j\/$
lives in a #3.\cr}

\halign{\indent #\hfil &\qquad #\hfil &\quad  #\hfil\cr

\state{w\phantom{{}^\prime}}{cottage}{mansion}
\state{z\phantom{{}^\prime}}{hovel}{palace} }

\medskip

\halign{\indent #\hfil &\qquad #\hfil &\quad  #\hfil\cr

\state{w^\prime}{cottage}{hovel} \state{z^\prime}{hovel}{cottage} } }
\bigskip

\noindent Suppose that $i\/$ were strictly to prefer state $w\/$ to
state $z\/$ and state $z^\prime\/$ to state $w^\prime\/$. The
interpretation of preferences as reflecting judgments about cardinal
differences in satisfaction of interests would suggest that $i\/$
considers a move from a hovel to a cottage to matter more for the
satisfaction of $j\/$'s interest than would a move from a mansion to a
palace. The argument for this implication is the same as before,
except now $i\/$'s contemplated sacrifice is a move from a cottage to
a hovel instead of from a palace to a mansion. If $i\/$ were to hold
all of the preferences that have been discussed in this paragraph and
in previous one, then the judgments-about-interests interpretation of
preferences would impute two inconsistent judgments to him. I will
demonstrate later in this paper (in lemma 10) that the assumptions
made in the preceding two sections are sufficient to rule out such
imputations of inconsistency. Technically, in addition to the
assumptions that have been introduced so far, lemma 10 also requires
an additional {\it double-cancellation condition} in the case of
two-agent communities and of some larger communitites. This condition
will be set forth when the lemma is stated formally.

Agent $j\/$ (or any other agent in the community) might either agree or
disagree with agent $i\/$'s assessments of how much difference various
changes of social state would make to the satisfaction of $j\/$'s interest.
The main result to be proved in this paper will be that, in the presence of
the assumptions that have already been made (along with the
double-cancellation condition), the additional assumption that agents
always agree with one another about such assessments is sufficient for the
liberal-succession relation and the unanimous-succession relation to
coincide.

\Section e{Equivalence of the welfare criteria: a preliminary result}

The model of a community set forth in sections \secref{b} and \secref{c} should
seem familiar in many respects, especially in relation to work of Debreu. As
in Debreu (1960), preferences of an individual agent are described in terms
of ordinal and topological assumptions along with an assumption regarding
separability. As in Debreu (1959), a welfare proposition regarding a
community of agents described in such terms is to be
investigated.\footnote{Debreu (1959) did not impose separability, however.} This
investigation divides logically into two parts. First a representation
theorem should be proved, asserting the existence of well behaved utility
functions which represent agents' preferences and also of analogous
functions which represent agents' interests. Then, taking advantage of these
numerical representations, the proposition about welfare should be proved.

Since welfare analysis provides the motivation for the representation
theorem, and also since the particular set of assumptions studied here in
this paper not be the only ones which yield the sort of representation that
is derived, I will begin by formulating the representation that is needed
and by proving that it implies that the liberal-succession and
unanimous-succession relations coincide.

Throughout this section and the remainder of the paper (except where
explicitly stated to the contrary), I will assume that conditions \eqn
a--\eqn f and \eqn j--\eqn m hold. The representation theorem that I will
prove later from these assumptions (and the two further ones discussed at
the end of the preceding section) contains the following three assertions
\eqn n--\eqn r regarding the possibility of representing agents' preferences
and interests by numerical functions.\footnote{The
scalar $\kappa_{i}$ in \eqn r could be eliminated if $p_{i}$ were adjusted by
subtracting it. However, the present formulation simplifies the proof of
lemma 5 below.}

\display n{v_{j} : X\rightarrow\Re \text{ is continuous and } \forall x\; \forall y\;
[xW_{j}y\iff v_{j}(x)\geq v_{j}(y)].}

\display o{p_{j}:X\rightarrow\Re \text{ is continuous and } \forall x\; \forall y\;
[xR_{j}y \iff p_{j}(x)\geq p_{j}(y)].}

\display r{\text{There exist scalars } \alpha_{ij} \text{ and
} \kappa_{i} \text{ such that } \forall i\; \forall x\; p_{i}(x)
= \Sigma_{j\,\leq\,n}\alpha_{ij}v_{j}(x) + \kappa_{i}.}

The first two of these assertions  are just the sort of ordinal
representations that one would expect. (Note, however, that \eqn n implies
the connectedness of interest relations, which has not been assumed directly
in \eqn a.) The real force of the representation is provided by assertion
\eqn r. What this assertion obviously accomplishes is to reflect numerically
the assumptions \eqn b and \eqn d that agents' preferences are based
on interests and separable in interests. The assertion states
significantly more than that, though. Each agent's preferences might
individually be based on interests and separable in interests, and yet
the numerical representation of someone's interest (agent $i\/$, say)
that entered one agent's (agent $j\/$) utility function might be an
arbitrary continuous, strictly increasing transformation of the
numerical representation of $i\/$'s interests that entered another
agent $k\/$'s utility function.  Assertion \eqn r states that the
representation of $i\/$'s interest that enters $j\/$'s utility
function must be a {\it strictly increasing affine transformation\/}
of the representation of $i\/$'s interest that enters $k\/$'s utility
function. This means that when the preferences of agents $j\/$ and
$k\/$ reflect {\it cardinal} comparisons of differences in
satisfaction of $i\/$'s interest (such as that it would make a greater
difference for $i\/$ to move from a hovel to a cottage than it would
make for him to move from the cottage to a mansion), then $j\/$ and
$k\/$ must agree with one another. As has already been explained in
the preceding section, the existence of such agreement throughout the
community regarding cardinal intrapersonal comparisons across social
states of each agent's interest is a key assumption of the proof that
liberal succession and unanimous succession coincide. This assumption
does not follow from the assumptions about the community that have
been made so far. An ordinal assumption that does imply assumption
\eqn r (in the presence of the other assumptions) will be introduced
below in \sec h.

Now all of the assumptions have been introduced that are needed to prove the
coincidence of the liberal-succession and unanimous-succession relations. In
the proof, it will be convenient to write \eqn r in matrix form. To do so,
define $v:X\rightarrow\Re^{n}$ and $p:X\rightarrow\Re^{n},$ where $n$ is the
number of agents in $i,$ by
\display p{\forall x\; \forall j\ [v(x)]_{j} = v_{j}(x)}
\noindent and
\display q{\forall x\; \forall j\; [p(x)]_{j} = p_{j}(x).}
\noindent Let $A$ be the matrix with coefficients $\alpha_{ij}.$ Then \eqn r
can be written in matrix notation as:
\display L{\forall x\; p(x) = Av(x) + \kappa.}

Now the proof of the welfare theorem will be presented in a series of
lemmas.

\lema a{$v(X)$ contains a non-empty open subset of
$\Re^{n}.$}

\begin{proof} Let $x_{j}V_{j}y_{j}$ for each $j,$ as is guaranteed  by \eqn k. By
continuity of $v_{j}$ and connectedness of $X,$ the interval
$[v_{j}(y_{j}),\,v_{j}(x_{j})]$ is a subset of $v_{j}(X).$ This interval has
nonempty interior, since $v_{j}$ represents $W_{j}.$ Since agents' interests
have product structure, then, $(v_{1}(y_{1}),\,v_{1}(x_{1}))\times \ldots
\times (v_{n}(y_{n}),\,v_{n}(x_{n}))\subseteq v(X).$\end{proof}

\lema b{$A$ is nonsingular.}

\begin{proof} By idiosyncracy of preferences, for each $i$ there exist $x$ and $y;$
such that $xP_{i}y$ and $\forall j\neq i\;xI_{j}y.$ Thus $p(x)-p(y) = A[v(x)
- v(y)]$ is a scalar multiple of the usual basis vector in dimension
$i$.\end{proof}

Let $B=(\beta_{ij}) = A^{-1}.$ Then, for every $i$ there exists a scalar
$\lambda_{i}$ such that

\display s{\forall x\;p_{i}(x) = \alpha_{ii}v_{i}(v) +
\sum_{j\,\neq\,i}\alpha_{ij}[\sum_{k \le n}\beta_{jk}p_{k}(x)] +
\lambda_{i}.}

Equation \eqn s provides an implicit characterization of the preferences of
agent $i$ in terms of $i$'s interest and other agents' preferences. However,
$p_{i}$ occurs on both sides of the equation. In order to have an explicit
characterization, $p_{i}$ should not occur on the right. That is, the term
$\sum_{j\,\neq\,i}\alpha_{ij}\beta_{ji}p_{i}(x)$ should be subtracted from
both sides of \eqn s, and then both sides of the resulting equation should
be divided by $1-\sum_{j\,\neq\,i}\alpha_{ij}\beta_{ji}$.   The result of
these operations, defining $\gamma_{ik} =
\sum_{j\,\neq\,i}\alpha_{ij}\beta_{jk},$
$\delta_{ii}=\alpha_{ii}\,/\,(1-\gamma_{ii}),$
$\mu_{i}=\lambda_{i}\,/\,(1-\gamma_{ii}),$ and
$\delta_{ik}=\gamma_{ik}\,/\,(1-\gamma_{ii})$ for $k\neq i,$ is

\display t{\forall x\;p_{i}(x) = \delta_{ii}v_{i}(x) +
\sum_{k\,\neq\,i}\delta_{ik}p_{k}(x) + \mu_{i}.}

There is one thing that has to be verified for this derivation of \eqn t to
be sound. That is that $\gamma_{ii}\neq 1.$ Otherwise, the final step of the
derivation would have been division by zero. This verification is now
provided.

Recall once again, that by idiosyncracy of $i$'s preference, there exist
states $x$ and $y$ such that $xP_{i}y$ and $\forall k\neq i\;xI_{k}y.$ By
nonpaternalism, $i$'s strict preference requires that $xV_{i}y.$ Equation
\eqn o implies that $p_{k}(x) = p_{k}(y)$ for all $k\neq i,$ so subtraction
of \eqn s from the corresponding equation for state $y$ yields
$[p_{i}(y)-p_{i}(x)] = $\hfill\break $\alpha_{ii}[v_{i}(y) - v_{i}(x)]\,
+\,\gamma_{ii}[p_{i}(y) - p_{i}(x)].$ By \eqn n, $[v_{i}(y) - v_{i}(x)]$ is
strictly positive. Therefore $\gamma_{ii}\neq 1$ if $\alpha_{ii}\neq 0.$
However, if $\alpha_{ii}=0,$ then it would follow from \eqn k, \eqn n, \eqn
o and \eqn r that $y_{j}R_{j}x_{j}$ (where $x_{j}$ and $y_{j}$ are the pair
of states that \eqn k guarantees to exist for $j),$ contrary to \eqn b.

Since $\gamma_{ii}\neq 1,$ then, preferences representable by \eqn r are
also representable by \eqn t. This latter representation is now studied
further.

\lema c{In \eqn t, for every $i$, $\delta_{ii} > 0.$}

\begin{proof} By idiosyncracy of preferences, there exist $x$ and $y$ such that
$xP_{i}y$ and $\forall j\neq i\;xI_{j}y.$ Thus, by nonpaternalism of
$i,\,xV_{i}y.$ Therefore $\delta_{ii} > 0$ follows from \eqn n, \eqn o and
\eqn t.\end{proof}

\lema d{In \eqn t, for every $i$ and $k,\;\delta_{ik}\geq 0.$}

\begin{proof} By lemma 3, it is sufficient to prove this inequality for $i\neq k.$  By
lemma 1, there exist a subset $Y\subseteq X$ and a nonempty open set
$U\subseteq\Re^{n}$ such that $v$ maps $Y$ onto $U.$ By lemma 2, $p(Y) =
A(U)$ is open in $\Re^{n}.$ For in define $q^{i}:Y\rightarrow\Re^{n}$ by
$[q^{i}(x)]_{j}=\delta_{ii}v_{i}(x)+\mu_{i}$ if $i=j,$ and
$[q^{i}(x)]_{j}=p_{j}(x)$ if $i\neq j.$ Note that, by \eqn t,
$q^{i}=Q^{i}p,$ where $Q^{i}$ is a matrix that will also be denoted by
$(\chi_{jk}).$ This matrix is the same as the identity matrix except for its
$i^{\rm th}$ row, and $\chi_{ii}=1$ and $\chi_{ik}= -\delta_{ik}$ for $k\neq
i.$ It is obvious (using row $i$ to expand the determinant) that $Q^{i}$ is
nonsingular, so $q^{i}(Y)=Q^{i}A(U),$ which is open in $\Re^{n}.$ Thus there
exist states $x$ and $y$ such that $q^{i}(x)-q^{i}(y)$ is a positive scalar
multiple of the usual basis vector in the $k$ dimension. That is,
$xE_{i}y,\;xP_{k}y,$ and $xI_{j}y$ for all other agents $j.$ By
nonpaternalism of $i,\;xR_{i}y,$ so $\delta_{ik}\geq 0$ follows from \eqn n,
\eqn o, and \eqn t.\end{proof}

These results have a useful geometric interpretation. Consider the vector
space that is obtained from the space of all continuous, real-valued
functions on $X$ when functions that differ by a constant are identified
with one another. Let $V$ be the finite-dimensional subspace generated by
$\{ v_{i}\}_{i\,\in\,I}.$ Lemma 1 asserts that $V$ is isomorphic to
$\Re^{n}.$ For a set $F$ of vectors in $V,$ let $K[F]$ be the convex cone
generated by $F.$ The possibility of an unambiguous improvement (i.e.,
assumption \eqn m) implies that $K[\{ v_{i}\}_{i\,\in\,I}\,\cup\,\{
p_{i}\}_{i\,\in\,I}]$ does not contain any linear subspace (except for
$\{\emptyset\})$ of $V.$ Lemma 2 asserts that $\{p_{i}\}_{i\,\in\,I}$ is a
basis of $V.$ Lemma 4 asserts that, for each  $i,\;p_{i}\in K[\{
v_{i}\}\,\cup\,\{p_{j}\}_{j\,\neq\,i}].$ These facts are now used to
establish the coincidence of the two social-choice relations.

\theor A{The additively-separable representation of
preferences discussed here is sufficient for the Mill's principle to be
equivalent to the Pareto principle. Specifically, suppose that the
environment is as described in \eqn a--\eqn f and \eqn j--\eqn m. If agents'
interests and preferences can be represented as in \eqn n---\eqn r, then
Pareto superiority and liberal succession coincide.}

\begin{proof} Since the Pareto relation is contained in the relation of liberal
succession, it is sufficient to prove that $x$ is Pareto superior to $y$ if
$x$ is a liberal succession of $y.$  A contradiction will be obtained from
the contrary assumption. In particular, assume that $J\subseteq I$ satisfies
\eqn i but that $yP_{k}x.$  Then  $p_{k}\in K[\{
v_{i}\}_{i\,\not\in\,J}\,\cup\,\{p_{j}\}_{j\,\in\,J}].$ Thus $K[\{
v_{i}\}_{i\,\not\in\,J}\,\cup\,\{p_{j}\}_{j\,\in\,I}]$ is not a subset of
$K[\{ v_{i}\}_{i\,\not\in\,J}\,\cup\,\{p_{j}\}_{j\,\in\,J}].$ By \eqn
m,\hfill\break $K[\{ v_{i}\}_{i\,\not\in\,J}\,\cup\,\{p_{j}\}_{j\,\in\,I}]$
contains no linear subspace (except $\{\emptyset\} )$ of $V.$ Therefore, by
\eqn l and Corollary 18.5.2 of Rockafellar (1970), there is an agent  $k$
such that $p_{k}\not\in K[\{
v_{i}\}_{i\,\not\in\,J}\,\cup\,\{p_{j}\}_{j\,\neq\,k}].$ (Note that \eqn l
rules out that $p_{j}$ could be a scalar multiple of $p_{k}$ for any $j\neq
k.)$ This contradicts lemma 4, since $k\not\in J$.\end{proof}

\Section f{Nonmalevolence}

Theorem 1 has antecedents in the work of Bergstrom ((1971), (1989), (1988)),
Green ((1979), (1982)), and Pearce (1983), all of which assume some version
of \eqn r. In the remaining part of this paper, it will be shown that the
assumptions on ordinal preference and interest relations that have been
introduced above (along with the double-cancellation condition and an
ordinal assumption that will imply agents' agreement about intrapersonal
interest comparisons) can replace \eqn r in the theorem. Before showing
this, though, I will now briefly discuss one difference between theorem 1
and it antecedents. The remainder of the paper is independent of this
discussion.

In the antecedent theorems just mentioned, the possibility of an unambiguous
improvement \eqn m has not been assumed but an additional assumption of {\it
nonmalevolence} has been made. Formally, agent $j$ is nonmalevolent if
\begin{itemize}
\item[(\eqnref{m}$'$)] $\forall x\; \forall y$ [if $\forall i\;xW_{i}y,$
then $xR_{i}y].$
\end{itemize}

Neither of assumptions \eqn m and $(13')$ implies the other. In particular,
note that the example in section 3 satisfies $(13')$ but not \eqn m. To see
that \eqn m does not imply $(13'),$ consider two agents and let $X$ be the
Euclidean plane, with $v_{1}(x) = x_{1}$ and $v_{2}(x) = x_{2}.$ Define
$p_{1}=2v_{1} - v_{2}$ and $p_{2}=2v_{2}-v_{1}.$ Obviously neither
preference relation defined by these utility functions satisfies $(13'),$
but $(1,\,1)$ is an unambiguous improvement over $(0,\,0).$

Conditional on the other assumptions that have been made here, though, the
assumptions \eqn m and $(13')$ are equivalent. It is easy to see that the
assumptions of this paper (specifically \eqn a, \eqn d, \eqn j, \eqn k and
$(13'))$ imply \eqn m. The converse implication is proved analogously to
theorem 1. Clearly $(13')$ implies that all of the coefficients
$\alpha_{ij}$ in \eqn r are nonnegative. Thus, if some agent $j$ does not
satisfy $(13'),$ then $p_{j}\not\in K[\{ v_{i}\}_{i\,\in\,I}].$ Thus $K[\{
v_{i}\}_{i\,\in\,I}\,\cup\,\{p_{j}\}_{j\,\in\,I}]$ is not a subset of $K[\{
v_{i}\}_{i\,\in\,I}].$ By \eqn m,
$K[\{v_{i}\}_{i\,\in\,I}\,\cup\,\{p_{j}\}_{j\,\in\,I}]$ contains only the
trivial linear subspace $\{\emptyset\}$ of $V,$ so there is an agent $k$
such that $p_{k}\not\in
K[\{v_{i}\}_{i\,\in\,I}\,\cup\,\{p_{j}\}_{j\,\neq\,k}].$ This contradicts
lemma 4, though, so the failure of $(13')$ for $j$ implies the failure of
\eqn m. That is, if \eqn m holds, then all agents must satisfy $(13').$

\Section g{Additive conjoint measurement of preference}

The proof of theorem 1 has relied heavily on two assumptions that are not
obvious consequences of what had previously been assumed. The first of these
is assumption \eqn n, that for each agent $j$ there is a continuous function
$v_{j}:X\rightarrow\Re$ such that $\forall x\;\forall y\; [xW_{j}y\iff
v_{j}(x)\geq v_{j}(y)].$ This assumption implies in particular that agents'
interest relations are connected on $X,$ which is a stronger connectedness
claim than the limited one that follows from assumptions \eqn a and \eqn b.
The other assumption is \eqn r, that the functions $v_{j}$ can be specified
in such a way that there exist scalars $\alpha_{ij}$ and $\kappa_{i}$ such
that $\forall i\; \forall x\;p_{i}(x) = \sum_{j\,\leq\,n}\alpha_{ij}v_{j}(x)
+ \kappa_{i}.$\footnote{\eqn n resembles \eqn m and \eqn q in its form, but it
follows routinely from previous assumptions, using a result of Debreu
(1959). The crucial difference is that $R_{i}$ is assumed to be connected,
while the connectedness of $W_{i}$ must be proved.}

These assumptions will be derived from ordinal and topological assumptions.
The first step to accomplish this is to formulate three conditions (i.e.,
$(21)$ - $(23)$ below) that jointly imply \eqn n and \eqn r. The first of
these conditions will be established in this section, and the second and
third in the next.

Call a function $f:X\rightarrow\Re$ {\it compatible with} an equivalence
relation $E$ on $X$ if, for all $x$ and $y,\;xEy$ implies that $f(x) =
f(y).$ Say that $f$ {\it is strictly increasing} with respect to a binary
relation $V$ on $X$ if, for all $x$ and $y,\;xVy$ implies that $f(x) >
f(y).$ Note that \eqn n is equivalent to the statement that, for each $j,$
there is a continuous function $v_{j}$ that is both compatible with $E_{j}$
and strictly increasing with respect to $V_{j}.$ The first of the three
conditions incorporates this information from \eqn n directly in a way that
closely resembles \eqn r. For all agents $i$ and $j,$ let $N(i)$ be a set of
agents, let $\nu_{i}$ be a scalar, and let $p_{i}$ and $u_{ij}$ be functions
from $X$ to $\Re .$ Condition \eqn r will be paraphrased by taking $N(i)$ to
be the set of $j$ such that $\alpha_{ij}\neq 0$ and $u_{ij}$ to be
$\alpha_{ij}v_{j}.$ Now the first two conditions can be stated:

\begin{enum}
\meti u{$\forall i\; \forall x\;p_{i}(x) = \sum_{_j\,\in\,N(i)}\;u_{ij}(x) +
\nu_{i},$ where each $p_{i}$ satisfies \eqn o and each $u_{ij}$ for  $j\in
N(i)$ is continuous, non-constant, and compatible with $E_{j},$}
\end{enum}

\noindent and

\begin{enum}
\meti v{Each $u_{ij}$ for $j\in N(i)$ is strictly increasing with respect to
$V_{j}.$}
\end{enum}

Conditions \eqn u and \eqn v do not quite imply \eqn r, because \eqn r
entails (according to the paraphrase that has just been described) the third
condition that\footnote{$\sigma_{hij}=\alpha_{hj}\,/\,\alpha_{ij}$ and
$\tau_{hij}=0.$}

\begin{enum}
\meti w{If $j\in N(h)\,\cap\,N(i),$ then there exist scalars $\sigma_{hij}$
and $\tau_{hij}$ such that $u_{hj}=\sigma_{hij}u_{ij} + \tau_{hij}.$}
\end{enum}

\lema e{Conditions \eqn u and \eqn w imply condition \eqn r.}

\begin{proof} Consider a well ordering of $I.$  For every $j,$ define $\iota (j)$ to
be the first $i$ such that $j\in N(i).$ Note that $\iota (j) \le j$ (because
$j\in N(j)$ by \eqn b, \eqn k, and \eqn u). For each $h,$ define $M(h)$ to
be the union of the sets $N(i)$ for $i$ preceding $h.$ For each $j,$ let
$v_{j}=u_{\iota (j)j}.$ Define $\alpha_{hj}=\sigma_{h\iota (j)j}$ if  $j\in
N(h)$ and $\iota (j) < h,\;\alpha_{hj}=1$ if $\iota (j)=h,$ and
$\alpha_{hj}=0$ otherwise. Define $\kappa_{h}=\nu_{h} +
\sum_{j\,\in\,M(h)\,\cap\,N(h)}\;\tau_{h\iota (j)j}.$ It is routinely
verified that these definitions satisfy \eqn r.\end{proof}

The remainder of this section is devoted to deriving \eqn u as a consequence
of ordinal and topological assumptions. Besides the assumptions that have
already been stated, only one further assumption is needed. This assumption,
called the {\it double-cancellation condition}, strengthens the separability
condition \eqn d in the case of an agent whose preferences depend only on
the interests of himself and one other agent:

\begin{enum}
\meti x{If $\exists i\neq j\; \forall x\;\forall y\;[xW_{i}y$ and $xW_{j}y$
jointly imply that $xR_{j}y],$ then the following condition holds for all
states $r,\,s,\,t,\,x,\,y,\,$ and $z$: if $rR_{j}x$ and $sR_{j}y$ and if
$rW_{i}t,\;xW_{i}s,\;yW_{i}z,\;rW_{j}y,\;sW_{j}t,$ and $xW_{j}z,$ then
$tR_{j}z.$}
\end{enum}

\noindent Like separability, this condition is necessary (even without any
other assumptions) for the existence of an additive conjoint representation.
Note that the condition is satisfied trivially if $N(j)$ is not a
two-element set.

Now it will be shown that the ordinal and topological assumptions that have
been made so far are sufficient to establish \eqn u. The proof that \eqn u
holds for agent $i$ depends on the cardinality of $N(i).$ If $N(i)$ has more
than one element, then \eqn u will follow from a representation theorem for
additive conjoint measurement due to Debreu (1959). If $N(i)$ is a singleton
(i.e., if $i$'s preferences depend only on his own interest), then it is
sufficient to take $p_{i}=u_{ii}=v_{i}$ and $\nu_{i}=0,$ where $v_{i}$
satisfies \eqn n. The existence of such a function  $v_{i}$ is now
established.

\lema f{Assumptions \eqn a, \eqn b, \eqn e, \eqn f, and
\eqn j imply assumption \eqn n, i.e., that for each agent $j$ there is a
continuous function $v_{j}:X\rightarrow\Re$ such that\\ $\forall x\;\forall
y\;[xW_{j}y\iff v_{j}(x)\geq v_{j}(y)].$}

\begin{proof} By a result of Debreu (1959), it is sufficient under these assumptions
to show that each $W_{j}$ is connected on $X.$ Consider any agent $j$ and
any pair of states $x$ and $y.$ By \eqn j, there exists a state $z$ such
that $xE_{j}z$ and $\forall i\neq j\;yE_{i}z.$ Either $yR_{j}z$ or $zR_{j}y$
by \eqn a, so either $yW_{j}z$ or $zW_{j}y$ respectively by \eqn b. Then by
\eqn a, either $yW_{j}x$ or $xW_{j}y$ respectively because
$xE_{j}z$.\end{proof}

In the case general that $R_{i}$ and $W_{i}$ may not coincide, the
derivation of \eqn u depends on Debreu's representation theorem, which is
now stated as lemma 7. For a proof, see Debreu (1959), or Krantz, et al.
(1971, \S 6.2, 6.11, 6.12). The continuity of $u_{ij}$ asserted in the lemma
is not explicitly mentioned in those sources, but it is clear from a careful
inspection of Debreu's proof, in which he has noted on pp. 22-24 that
$u_{ij}$ is obtained from a continuous function and that the several steps
of the construction of $u_{ij}$ preserve continuity.\footnote{Unlike Debreu's
proof, the proof given by Krantz, et al. does not assume the separability of
$X,$ and it does not assert the continuity of the functions $u_{ij}.$ I do
not know whether the functions constructed in that proof actually are
continuous. Since continuity of the functions representing agents' interest
relations was used in lemma 1 and will be used again in the next section, I
have asserted the separability of $X$ in \eqn f.}

\begin{lemma}[Representation theorem for additive conjoint
measurement] Suppose that $R_{i}\neq W_{i}$ and that assumptions \eqn a -
\eqn f, \eqn k and \eqn x are satisfied. Suppose that
$X=\Pi_{j\,\in\,I}\;X_{j},$ where each $X_{j}$ is a topological space and
$X$ has the product topology, and that, for each $j$, $\forall x\;\forall
y\;[x_{j}=y_{j}$ implies $xE_{j}y].$ Then $\forall x\;p_{i}(x) =
\sum_{j\,\in\,N(i)}u_{ij}(x) + \nu_{i},$ where $i\in N(i)\subseteq
I,\;p_{i}$ satisfies \eqn o and each $u_{ij}$ for $j\in N(i)$ is continuous
and non-constant and factors through $X_{j}$ (i.e., $U_{ij}(x)$ depends only
on $x_{j}).$ \label{lem:g} \end{lemma}

Lemma 7 falls short of the objective of this section in two respects. First,
even given the hypotheses of the theorem, the conclusion that each function
$u_{ij}$ factors through $X_{j}$ is weaker than the desired conclusion that
each $u_{ij}$ is compatible with $W_{j}.$  Second, the hypothesis that $X$
is a cartesian product of spaces that individually determine agents'
interest levels is stronger than the product-structure assumption \eqn j.
The cartesian-product representation restricts the economic applicability of
the lemma to private-goods environments, since the level of provision of a
public good would be a coordinate of the social state that would affect the
interests of all agents. The conclusion of lemma 7 can actually be
strengthened to include compatibility of each $u_{ij}$ with $E_{j},$ and the
hypothesis of a cartesian-product representation can be weakened to
assumption \eqn j if \eqn g is also assumed. These amendments are now made
in lemma 8 and lemma 10.

\lema h{Suppose that  assumptions \eqn a - \eqn f, \eqn k
and \eqn x are satisfied. Suppose that $X=\Pi_{j\,\in\,I}X_{j},$ where each
$X_{j}$ is a topological space and $X$ has the product topology, and that,
for each $j,\;\forall x\;\forall y\;[x_{j}= y_{j}$ implies $xE_{j}y].$ Then
$\forall x\;p_{i}(x) = \sum_{j\,\in\,N(i)}u_{ij}(x) + \nu_{i},$ where $i\in
N(i)\subseteq I,\;p_{i}$ satisfies \eqn o and each $u_{ij}$ for $j\in N(i)$
is continuous, non-constant, and compatible with $E_{j}.$}

\begin{proof} Lemma 7 establishes everything except for the compatibility of $u_{ij}$
with $E_{j},$ and it establishes that each $u_{ij}$ factors through $X_{j}.$
A contradiction will be derived from the assumption that, for some $j\in
N(i),\;u_{ij}$ is not compatible with $E_{j}.$ Specifically, suppose that
$xE_{j}y$ but that $u_{ij}(x) < u_{ij}(y).$ Let $z_{j}=x_{j}$ but let
$z_{h}=y_{h}$ for all $h\neq j.$ Then $\forall h\;zE_{h}y,$ so $zR_{i}y$ by
\eqn b. However, $p_{i}(z) < p_{i}(y),$ contradicting  \eqn o.\end{proof}

A technical result is needed in order to relax the cartesian-product
hypothesis to \eqn j. Define a set $Y$ of states to be {\it saturated} with
respect to an equivalence relation $E$ if $x\in Y$ whenever, for some
$y,\;y\in Y$ and $xEy.$ Note that a function is compatible with $E$ if and
only if its level sets are saturated with respect to $E.$ It will be
required that the property of saturation with respect to the relations
$E_{j}$ should be preserved when topological closures are taken. Lemma 9
asserts that this requirement can be met.

\lema i{If \eqn a, \eqn f, and \eqn g hold and $Y$ is a
set of states that is saturated with respect to $E_{j},$ then $c\ell (Y)$ is
saturated with respect to $E_{j}.$}

\begin{proof} Equation \eqn a entails that $E_{j}$ is an equivalence relation. By \eqn f, the
graph of $E_{j}$ is closed. Suppose that $c\ell (Y)$ is not saturated with
respect to $E_{j}.$ Then there exist states $y\in c\ell (Y)$ and $x\not\in
 c\ell (Y)$ such that $xE_{j}y.$ Since $x\not\in c\ell (Y),$ there is an open
set $U$ containing $x$ and disjoint from $Y.$ By \eqn g, there is an open
set $V$ containing $y$ and satisfying $\forall v\in V\;\exists u\in
U\;uE_{j}v.$ Thus $Y$ cannot be saturated with respect to $E_{j},$ because
$Y\,\cap\,V\neq\emptyset$.\end{proof}

Recall that if $E$ is an equivalence relation on $X,$ then the {\it quotient
space} $X\,/\,E$ is the set of $E$-equivalence classes of elements of $X,$
with the finest topology that makes the mapping from elements to their
equivalence classes continuous.

\lema j{Suppose that  assumptions \eqn a--\eqn g, \eqn j,
\eqn k and \eqn x are satisfied. Then $\forall y\;p_{i}(x) =
\sum_{j\,\in\,N(i)} u_{ij}(x) + \nu_{i},$ where $i\in N(i)\subseteq
I,\;p_{i}$ satisfies \eqn o and each $u_{ij}$ for $j\in N(i)$ is continuous,
non-constant, and compatible with $E_{j}.$  That is, \eqn u is satisfied.}

\begin{proof} Begin by defining $xE_{I}y$ if and only if $\forall i\;xE_{i}y.$
Obviously $E_{i}$ is an equivalence relation and each $W_{i}$ induces a
relation on $X\,/\,E_{I}.$ By \eqn b, each $R_{i}$ also induces a relation
on $X\,/\,E_{I}.$ (`$W_{i}$' and `$R_{i}$' will denote the relations on
$X\,/\,E_{I},$ as well as the relations on $X).$ There are functions $v_{i}$
and $p_{i}$ satisfying \eqn n and \eqn o by lemma 6 and (Debreu (1959)),
respectively, and these induce functions from $X\,/\,E_{I}$ to $\Re$ that
are compatible with $E_{i}$ (resp. $I_{i})$ and strictly increasing with
respect to $V_{i}$ (resp. $P_{i}).$ The induced functions are continuous by
Bourbaki (1965, I.3.4, Prop. 6).  Thus the induced functions satisfy \eqn n
and \eqn o relative to the induced relations on $X\,/\,E_{I}.$  Therefore
the relations $R_{i}$ and $W_{i}$ on $X\,/\,E_{i}$ have closed graph. Now,
for each $i,$ define $X_{i}=X\,/\,E_{i}.$ There is a canonical bijection
between $X\,/\,E_{I}$ and $\Pi_{j\,\in\,I}X_{j}.$ By lemma 9 and Bourbaki
(1965, I.5.4, Prop 6 and corollary to Prop. 7), this bijection is a
homeomorphism. Thus $R_{i}$ and $W_{i}$ induce relations with closed graph
on $\Pi_{j\,\in\,I}X_{j}.$ These induced relations also satisfy the other
hypotheses of lemma 8. The conclusion of this lemma is derived by composing
the canonical surjection of $X$ onto $\Pi_{j\,\in\,I}X_{j}$ with the
functions from $\Pi_{j\,\in\,I}X_{j}$ to $\Re$ that lemma 8 asserts to
exist.\end{proof}

\Section h{Publicity of cardinal intrapersonal interest comparisons}

Conditions \eqn v and \eqn w still have to derived. In combination
with \eqn u, condition \eqn v states that, when one agent cares about
the interest of another at all, then he strictly prefers an increase
in the interest of that agent if the interests of others are held
constant. Condition \eqn w states that agents' preferences reflect
agreement about the cardinal magnitudes of intrapersonal interest
differences between states. (Note that this as an assertion of
coincidence between various agents' subjective preferences regarding
relevant sets of social states. It is not being asserted that there is
any objective basis for making interpersonal comparisons of either
preferences or interests.) Thus, both of these conditions are
substantive ones.

The assumptions that have been made so far are insufficient to guarantee
either of these conditions. However, both conditions can be derived if the
assumption $(27)$ that will be stated below is added to those already made.
The assumption states (in the context of \eqn d) that agents' preference
relations  reflect agreement about the directions and relative magnitudes of
pairs of interest differences, so evidently it is necessary for \eqn u, \eqn
v, and \eqn w to hold. The statement of $(27)$ will be simplified by the
following lemma.

\lema k{If assumptions \eqn b, \eqn j, \eqn k, and \eqn u
hold, then $\forall i\,\;i\in N(i)$ and
\display y{j\in N(i)\iff \exists x\;\exists y\;[xP_{i}y \text{ and } \forall h\neq
j\;xE_{h}y].}}

\begin{proof} If $j\not\in N(i)$ and  $\forall h\neq j\;xE_{h}y,$ then \eqn u implies
that $xI_{i}y.$ Thus, if $xP_{i}y$ and $\forall h\neq j\;xE_{h}y,$ then
$j\in N(i).$ By \eqn b and \eqn k, this implication entails that  $i\in
N(i).$ The converse implication follows from \eqn o and the nonconstancy and
compatibility assertions of \eqn u. If $j\in N(i),$ then there exist states
$w$ and $z$ such that $u_{ij}(w) > u_{ij}(z).$ By \eqn j, there are states
$x$ and $y$ such that $xE_{j}w,\;yE_{j}z,$ and $\forall h\neq j\;xE_{h}y.$
These states therefore satisfy $xP_{i}y,$ as well.\end{proof}

Now the ordinal assumption regarding publicity of cardinal intrapersonal
interest comparisons can be stated. For brevity, this assumption will be
called {\it interest cardinality.}

\begin{enum}
\meti z{Let $N(i)$ be defined by \eqn y. Suppose that $j\in
N(h)\,\cap\,N(i).$ Then the following implication holds for all states
$w_{h},\,w_{i},\,x_{h},\,x_{i},\,y_{h},\,y_{i},\,z_{h},$ and $z_{i}$
satisfying $w_{h}E_{j}w_{i},\,$ $\,x_{j}E_{j}x_{i},\,$ $y_{h}E_{j}y_{i},\,$
$z_{h}E_{j}z_{i},$ and $\forall g\neq j\;[w_{h}E_{g}x_{h},\,$ $
w_{i}E_{g}x_{i},\,$ $y_{h}E_{g}z_{h}$ and $y_{i}E_{g}z_{i}]$: if
$w_{h}R_{h}y_{h},\,$ $z_{h}R_{h}x_{h},$ and $y_{i}R_{i}w_{i},$ then
$z_{i}R_{i}x_{i}.$}
\end{enum}

To understand this assumption, think of agents $h$ and  $i$ as forming their
preferences by weighing interest gains to agent $j$ against interest losses
to other agents. Agent $h$ judges that $j$'s gain moving from $w_{h}$ to
$y_{h}$ would not outweigh others' losses but that $j$'s gain moving from
$x_{h}$ to $z_{h}$ would at least balance others' losses. Since others'
interest is the same in $w_{h}$ as in $x_{h}$ and also in $y_{j}$ as in
$z_{h},\;h$ must judge that $j$ gains at least as much by moving from
$x_{h}$ to $z_{h}$ as by moving from $w_{h}$ to $y_{h}.$ If $i$ shares this
judgment, then he should also judge that $j$ gains at least as much by
moving from $x_{i}$ to $z_{i}$ as by moving from $w_{i}$ to $y_{i},$ since
$j$ has the same interest in each of the `$i$' states as in the
corresponding `$h$' state. If $i$ also judges that $j$'s gain moving from
$w_{i}$ to $y_{i}$ is sufficient to balance others' losses, then he should
judge as well that $j$'s gain in moving from $x_{i}$ to $z_{i}$ is
sufficient to balance others' losses, as the conclusion of \eqn z reflects.

\lema l{If \eqn b, \eqn j, \eqn k, \eqn u and \eqn z hold
and $j\in N(i),$ then $u_{ij}$ is strictly increasing in $V_{j}.$ That is,
under these hypotheses \eqn v holds and $v_{j}=u_{ij}$ satisfies \eqn n for
$j.$}

\begin{proof} First, in \eqn z substitute $w$ for $w_{h}$ and $w_{i},\;x$ for $x_{h}$
and $x_{i},\;y$ for $y_{h}$ and $y_{i},$ and $z$ for $z_{h}$ and $z_{i}.$
Next substitute $j$ for $i,\;i$ for $h,$ and $x$ for $w$ and $y.$ The
resulting statement is the implication that, if $zR_{i}x$ and $\forall g\neq
j\;xE_{g}z,$ then $zR_{j}x.$ Now, consider any states $x$ and $w$ such that
$xV_{j}w.$ By \eqn j there exists $z$ such that $wE_{j}z$ and $\forall g\neq
j\;xE_{g}z.$ Note that $xV_{j}z,$ so that $xP_{j}z$ by \eqn b and therefore
$xP_{i}z$ by \eqn z (using the substitutions that have just been made).
Therefore by \eqn u, $u_{ij}(x) > u_{ij}(z) = u_{ij}(w)$.\end{proof}

Now the affine-relation condition \eqn w can be studied. The first step is
to derive from \eqn w a condition that resembles \eqn z more closely.

\lema m{If \eqn b, \eqn j, \eqn k, \eqn u and \eqn z hold
and $j\in N(h)\,\cap\, N(i),$ but \eqn w does not hold, then there exist
states $x,\,y,$ and $z$ such that $xV_{j}y,\;yV_{j}z,$ and
\display A{[u_{hj}(y)-y_{hj}(z)]\,/\,[u_{hj}(x) - u_{hj}(z)]\neq
[u_{ij}(y)-u_{ij}(z)]\,/\,[u_{ij}(x) - u_{ij}(z)].}}

\begin{proof} Consider any states $r$ and $s$ such that $rV_{j}s.$  Define a function
$f:X\rightarrow\Re$ by $$f(t)=u_{ij}(s) + \{ [u_{ij}(r) -
u_{ij}(s)]\,/\,[u_{hj}(r) - u_{hj}(s)]\}\;[u_{hj}(t) - u_{hj}(s)].$$
\noindent Note that the denominator of the fraction in the expression is
nonzero by lemma 12, and that $f(t)=u_{ij}(t)$ for $t=r,\,s.$ If \eqn w does
not hold, then there must be some other state $t$ such that $f(t)\neq
u_{ij}(t),$ and it is impossible that $tE_{j}r$ or $tE_{j}s.$ Let $x,\,y,$
and $z$ be $r,\,s,$ and $t$ arranged in order of decreasing interest
afforded to $j.$ The conclusion of the lemma is obtained by noting that
$$[u_{hj}(y)-u_{hj}(z)]\,/\,[u_{hj}(x)-u_{hj}(z)]=
[f(y)-f(z)]\,/\,[f(x)-f(z)],$$ that $u_{ij}\;(t)\neq f(t)$ for exactly one
of the values $t=x,\,y,\,z,$ and that $[a-c]\,/\,[b-c]$ is strictly monotone
in each of its variables on $\{(a,\,b,\,c)\,| \,a>b>c\}$.\end{proof}

\lema n{Suppose that \eqn b, \eqn f, \eqn j, \eqn k and
\eqn z hold that $j\in N(h)\cap N(i),$ where $h$ and $i$ are distinct
agents. Suppose that each of $N(h)$ and $N(i)$ contains an element distinct
from $j.$ Then \eqn w holds with respect to $j.$}

\begin{proof} Consider any three states $x,\,y,$ and $z$ such that $xV_{j}y$ and
$yV_{j}z.$ To simplify notation, it will be assumed without loss of
generality that $j$ is neither $h$ nor $i.$\footnote{This entails that $h$ and
$i$ are elements of $N(h)$ and $N(i),$ respectively, that are distinct from
$j.$ In general, agents $h'\in N(h)$ and  $i'\in N(i)$ will have to be
specified that are distinct from $j.$} Let $r_{h}V_{h}s_{h}$ and
$r_{i}V_{i}s_{i},$ as guaranteed by \eqn k. Choose any natural number $n$
sufficiently large so that

\display B{\begin{gathered} u_{hj}(x)-u_{hj}(z) <
n\,[u_{hh}(r_{h})-u_{hh}(s_{h})] \text{ and }\\
u_{ij}(x) - u_{ij}(z) < n\;[u_{ii}(r_{i})-u_{ii}(s_{i})].\end{gathered}}

\noindent Because the functions $u_{hh}$ and $u_{ii}$ are continuous and $X$
is connected, there exist $t_{h}$ and $t_{i}$ such that
$u_{hj}(x)\,-\,u_{hj}(z)=n\;[u_{hh}(t_{h})\,-\,u_{hh}(s_{h})]$ and
$u_{ij}(x)\,-\,u_{ij}(z)=$\hfill\break $n\;[u_{ii}(t_{i})-u_{ii}(s_{i})].$
Now, because $u_{hj}$ is continuous and $X$ is connected, there exists a
function $\eta_{n}:\{0,\,1,\ldots ,\,n\}\rightarrow X$ such that, for $m\leq
n,\,u_{hj}(\eta_{n}(m))=u_{hj}(z)\,+\,
m\;[u_{hj}(t_{h})\,-\,u_{hj}(s_{h})].$ (In particular, this implies that
$\eta_{n}(0)E_{j}z$ and $\eta_{n}(n)E_{j}x.)$ Assumption \eqn z will now be
used to show that, for all $m\leq
n,\,u_{ij}(\eta_{n}(m))=u_{ij}(z)\,+\,m\;[u_{ij}(t_{i})\,-\,u_{ij}(s_{i})].$
Since $\eta_{n}(0)E_{j}z$ and $\eta_{n}(n)E_{j}x,$ this is equivalent
to\hfill\break $\forall m<n\;u_{ij}(\eta_{n}(m+1)) \,-\,
u_{ij}(\eta_{n}(m))=u_{ij}(t_{i})\,-\,u_{ij}(s_{i}).$ Suppose, to the
contrary, that $u_{ij}(\eta_{n}(m+1))\,-\,u_{ij}(\eta_{n}(m))\neq
u_{ij}(t_{i})\,-\,u_{ij}(s_{i})$ for some $m<n.$ Without loss of generality,
suppose that $u_{ij}(\eta_{n}(m+1))\,-\,u_{ij}(\eta_{n}(m)) <
u_{ij}(t_{i})\,-\,u_{ij}(s_{i}).$ Then, for some $p,$
$u_{ij}(\eta_{n}(p+1))-u_{ij}(\eta_{n}(p))>u_{ij}(t_{i})-u_{ij}(s_{i}).$
(Otherwise $u_{ij}(z)-u_{ij}(x)\,=\,$
$\sum_{k\,<\,n}\;[u_{ij}(\eta_{n}(k+1))-u_{ij}(\eta_{n}(k))]<
n\;[u_{ij}(t_{i}-u_{ij}(s_{i})].)$ Now, using \eqn j, states
$w_{h},\,w_{i},\,x_{h},\,x_{i},\,y_{h},\,y_{i},\,z_{h},$ and $z_{i}$ will be
specified that satisfy the hypotheses of \eqn z. Here, when a state is
subscripted by `$g$', the subscript takes both values $h$ and $i.$ The
subscript `$f$' takes all values except $j$ and the value taken by $g$ in
the same expression. (That is, `$f$' can always take any value except
$h,\,i,$ and $j,$ and it can take whichever of $\{h,\,i\}$ is not taken by
$g.)$ Let $q$ be a fixed, arbitrary state.

Suppose that $w_{g}E_{j}\eta_{n}(p),$ $x_{g}E_{j}\eta_{n}(m),$
$y_{g}E_{j}\eta_{n}(p+1),$ and $z_{g}E_{j}\eta_{n}(m+1);$ that
$w_{g}E_{g}t_{g},$ $x_{g}E_{g}t_{g},$ $y_{g}E_{g}s_{g},$ and
$z_{g}E_{g}s_{g};$ and that $w_{g}E_{f}q,$ $x_{g}E_{f}q,$ $y_{g}E_{f}q,$ and
$z_{g}E_{f}q.$

With respect to the states so specified, \eqn z, fails to hold if \eqn u
holds. Since this contradicts the hypothesis of the lemma,  $\forall m\leq
n\;u_{ij}(\eta_{n}(m))=u_{ij}(z) + (m\,/\,n)\;[u_{ij}(x) - u_{ij}(z)].$

For any $n$ large enough to satisfy \eqn B, there is a natural number
$\mu_{n} < n$ satisfying $\eta_{n}(\mu_{n}+1)V_{j}y$ and
$yW_{j}\eta_{n}(\mu_{n}).$ Thus both  $[u_{hj}(y) -
u_{hj}(z)]\,/\,[u_{hj}(x) - u_{hj}(z)]$ and $[u_{ij}(y) \,-\,
u_{ij}(z)]\,/\,[u_{ij}(x) \,-\, u_{ij}(z)]$ are in the interval
$[\mu_{n}\,/\,n,\;(u_{n}\,+\,1)\,/\,n).$ Since $n$ can be arbitrarily large,
$[u_{hj}(y)\,-\,u_{hj}(z)]\,/\,[u_{hj}(x) \,-\, u_{hj}(z)]\,=\,$ $[u_{ij}(y)
\,-\, u_{ij}(z)]\,/\,[u_{ij}(x) \,-\, u_{ij}(z)].$ Therefore \eqn w holds,
by lemma 13.\end{proof}

\lema o{If conditions \eqn b, \eqn f, \eqn j, \eqn k,
\eqn u and \eqn z hold, then condition \eqn w holds for all agents $h,\;i,$
and $j.$}

\begin{proof} In view of lemma 14, \eqn w only has to be established now in the case
that $h=j,\;N(j) = \{ j\}$ and $j\in N(i)$ for some $i\neq j.$ In that case,
simply take $u_{hj}=u_{ij}$ for some $i$ satisfying $j\in N(i)$.\end{proof}

The foregoing lemmas immediately establish the following theorem.

\theor B{Suppose that  assumptions \eqn a--\eqn f, \eqn
j--\eqn m, \eqn x and \eqn z are satisfied, and that either \eqn g
is satisfied or else $X$ possesses the cartesian-product structure described
in lemma 7. Then Pareto superiority and liberal succession coincide.}

\Section i{An alternative cardinality condition on preferences}

Within many communities there is wide, if approximate agreement, regarding
cardinal intrapersonal comparisons of interest. It is uncontroversial that
someone's interest would be more significantly advanced if he were to move
from a hovel to a decent house than if he were to move from the house to a
mansion, for example. The prevalence of this kind of shared intuition about
welfare contributes to the plausibility of condition \eqn z.

However, some people may find it less plausible that there is public
agreement about intrapersonal cardinal interest comparisons, than that there
is public agreement about intrapersonal cardinal preference
comparisons.\footnote{I am indebted to Arthur Robson for stating the two
considerations that I discuss now.} For one thing, it is often more obvious
what people prefer than what is objectively good for them. For another,
people who endorse expected-utility theory are already committed to accept
one form of the cardinality of persons' preferences, even if they do not
accept the cardinality of persons' interests. People who hold these views
may prefer to assume the publicity of intrapersonal cardinal preference
comparisons rather than to assume \eqn z. It will be shown here that such an
assumption can be formulated in a way similar to \eqn z, and that the formal
assumption (together with the other axioms) implies the existence of common
additive interest factors.

Consider how to express the agreement of two agents, $j$ and $k,$ about
intrapersonal cardinal preference comparisons concerning agent $i.$ Suppose
that $\forall h\not\in \{i,\,j\}\;[wI_{h}y$ and $xI_{h}z],$ and that
$wE_{j}y$ and $xE_{j}z.$ Make an analogous assumption with $k$ substituted
for $j,$ but with respect to a different set of social states. Specifically,
suppose that $\forall h\not\in \{i,\,k\}$ $[w^\prime I_{h}y^\prime$ and
$x^\prime I_{h}z^\prime],$ and that $w^\prime E_{j}y^\prime$ and $x^\prime
E_{j}z^\prime.$ Let $i$ be indifferent between the corresponding social
states mentioned in these two assumptions. That is, let $wI_{i}w^\prime,$
$xI_{i}x^\prime,$ $yI_{i}y^\prime,$ and $zI_{i}z^\prime.$

Now suppose that $xR_{j}w,\;yR_{j}z,$ and $w^\prime R_{k}x^\prime.$ By
reasoning analogous to that regarding \eqn z, the two preferences of $j$
imply that $j$ must regard $i$'s interest gain from being in $x$ rather than
$w$ as being at least as great as $i$'s interest gain from being in $z$
rather than $y$ would be. The preference of $k$ implies that $k$ must regard
$i$'s interest gain from being in $x^\prime$ rather than $w^\prime$ as being
insufficient to outweigh strictly the net disadvantages of $x^\prime$
relative to $w^\prime$ for everyone else. (Note that these will be in terms
of preference for everyone except $k,$ and in terms of interest for $k.)$
Thus, if $k$ agrees with $j$ about intrapersonal cardinal preference
comparisons regarding $i,$ then $k$ should also regard $i$'s interest gain
from being in $z^\prime$ rather than $y^\prime$ as being insufficient to
outweigh strictly the net disadvantages of $z^\prime$ relative to $y^\prime$
for everyone else. That is, it should be the case that $y^\prime
R_{k}z^\prime.$

The implication from these various assumptions to their conclusion jointly
constitute the assumption that {\it intrapersonal cardinal preference
comparisons are public:}

\begin{enum}
\meti C{Define $j\in S(i)$ if and only if, for some states $x$ and $y,$
$xE_{i}y,$ $\forall h\not\in \{ i,\,j\}$ $xI_{h}y,$ and $xP_{h}y.$ If $j\in
S(h)\,\cap\, S(i),$ then following implications holds for all states
$w_{h},\,w_{i},\,x_{h},\,x_{i},\,y_{h},\,y_{i},\,z_{h},$ and $z_{i}$
satisfying $w_{h}I_{j}w_{i},$ $x_{h}I_{j}x_{i},$ $y_{h}I_{j}y_{i},$
$z_{h}I_{j}z_{i},$ $\forall g\not\in \{ h,\,j\}$ $[w_{h}I_{g}x_{h}$ and
$y_{h}I_{g}z_{h}],$ $\forall g\not\in \{ i,\,j\}$ $[w_{i}I_{g}x_{i}$ and
$y_{i}I_{g}z_{i}],$ $w_{h}E_{h}x_{h},$ $y_{h}E_{h}z_{h},$ $w_{i}E_{i}x_{i},$
and $y_{i}E_{i}z_{i}$: if $w_{h}R_{h}y_{h},$ $z_{h}R_{h}x_{h},$ and
$y_{i}R_{i}w_{i},$ then $z_{i}R_{i}x_{i}.$}
\end{enum}

It seems very plausible that this condition can play an analogous role to
\eqn z in guaranteeing the coincidence of Pareto superiority and liberal
succession on the basis of ordinal and topological assumptions.

\Section j{Coincidence without separability or publicity}

Theorem 2 and the alternative just suggested each hypothesize (a) product
structure of interests and separability of preferences in interests, and (b)
public agreement regarding some form of intrapersonal cardinal comparison.
The standard construction of nonpaternalistic preferences, by beginning with
cardinal interest-representation functions as in theorem 1, yields a family
of preferences that satisfy all of these qualitative hypotheses. Although
the example provided in section 3 has shown that the hypotheses of
interest-determination and nonpaternalism alone are too broad to
characterize the class of nonunanimous preference profiles on which the
liberal principle and unanimity coincide, it might well be suspected that
the additional qualitative hypotheses would prove to be necessary as well as
sufficient for this coincidence. A counterexample that disproves this
conjecture is now provided.\footnote{This example will involve preferences that
are represented by utility functions defined by taking the minimum of
several numbers. Such utility functions are pointwise limits of sequences of
CES utility functions. Every CES utility function represents a preference
relation satisfying the separability assumption. Thus, despite the example
there may still be a close connection between separability and the
coincidence of the two welfare relations.}

Let there be three agents,  $I=\{1,\,2,\,3\},$ and let $X=\Re_{+}^{3}.$
Define

\display D{v_{i}(x) = x_{i},}

\noindent and define

\display E{p_{i}(x) =\min\;\{x_{i}/2,\;x_{j},\;x_{k}\} \text{, where }
I=\{i,\,j,\,k\}.}

\noindent Then define $W_{i}$ and $R_{i}$ as in \eqn n and \eqn o.

It is evident from \eqn D and \eqn E that agents' preferences are determined
by interests, and that the agents are not unanimous. It will now be shown
that their preferences are also nonpaternalistic and that the relation of
liberal succession coincides with that of Pareto improvement. Since the
latter claim implies the former, only the coincidence of the two welfare
orderings needs to be shown. Recall that every Pareto improvement is
automatically a liberal successor, so it is sufficient to establish the
converse inclusion.

Suppose, then, that $y$ is a liberal successor of $x.$ Specifically, suppose
that

\display F{\forall i\in C\;p_{i}(y)\geq p_{i}(x),}

\noindent that

\display G{\forall j\not\in C\;y\geq x,}

\noindent and that

\display H{p_{k}(x) > p_{k}(y).}

\noindent By \eqn H, $y$ is not a Pareto improvement over $x.$ This will be
shown to be a contradiction.

By \eqn E and \eqn H,

\display I{\min\;\{ y_{i},\,y_{j},\, {y_{k}}/{2}\}
<\min\;\{x_{i},\,x_{j},\,{x_{k}}/{2}\}.}

\noindent By \eqn E and \eqn H,$k \not\in C,$ so  $y_{k}\geq x_{k}$ by \eqn
G. Therefore \eqn I implies that, for one of the other agents $h\in \{i,\,j\},$

\display J{y_{h}=\min\;\{y_{i},\,y_{j},\,{y_{k}}/{2}\} <
\min\;\{x_{i},\,x_{j},\,{x_{k}}/{2}\}\leq x_{h}.}

\noindent Therefore $h\in C$ by \eqn G. But, by \eqn E and \eqn J,

\display K{p_{h}(y)\leq {y_{h}}/{2}
< \min\;\{{x_{h}}/{2},\,x_{i},\,x_{j},\,x_{k}\} = p_{h}(x).}

\noindent Therefore $h\not\in C$ by \eqn F, so a contradiction has been
reached. This establishes that liberal succession coincides with Pareto
improvement, and consequently that all agents have nonpaternalistic
preferences.

However, agents' preferences are not separable in interests, and cardinal
interpersonal comparisons of neither preferences nor interests are public.
To see that the preferences of agent 1 are inseparable, let $J=\{1\}$ and
$K=\{2,\,3\},$ and define $w=(1,\,2,\,2),$ $x=(2,\,2,\,2),$ $y=(1,\,0,\,0),$
and $z=(2,\,0,\,0).$ Then \eqn d fails to hold because, although its
hypothesis holds, $xP_{1}w$ while $yR_{1}z.$ By the symmetry of the example,
the preferences of the other two agents are not separable in interests
either.

To show that intrapersonal cardinal interest comparisons are not public, let
$i=1$ and $j=2$ in \eqn z. Define $w=z=(3,\,2,\,2)$ and $x=y=(2,\,2,\,2).$
Then the hypothesis of \eqn z holds, but $wR_{1}x$ and not $yR_{1}z,$ so the
conclusion of \eqn z fails to hold. Therefore \eqn z (which is an
implication) fails to hold.

Intrapersonal cardinal preference comparisons can similarly be shown not to
be public. To do so, let $i=1,\;j=2,$ and $k=3$ in \eqn C, and define:
$w=z=(3,\,3,\,3)$, $x=y=(2,\,3,\,3)$,
$w^\prime=z^\prime=(3,\,3,\,7)$, and $x^\prime=y^\prime=(2,\,3,\,7)$.
These social states have the feature that agents 1 and 2 always
have ``selfish'' preferences among them (that is, the preference and
interest relations of each agent among these states coincide), and that
agent 3 has ``selfish'' preferences among $\{w,\,x,\,y,\,z\}$ but has
totally ``altruistic'' preferences for the interest of agent 1 among
$\{w^\prime,\,x^\prime,\,y^\prime,\,z^\prime\}.$ From these considerations,
it can easily be seen that \eqn C fails to hold.

\Section k{Examples and conclusion}

Strictly speaking, Mill's principle concerns only the permissibility of
private activities. Contemporary libertarians advocate that public
authorities should not have the power to take any action that Mill's
principle would prohibit to the coalition of persons who prefer the action.
Such a position obviously has strong distributive implications. An
alternative view would hold that public authorities can legitimately take
some actions that would be prohibited to private persons and coalitions, but
would also recognize that public actions should be structured in a way that
will minimize ``transactions costs.'' On this view, the public authorities
should not take an action if some person or coalition can propose an
alternative action that the liberal principle would endorse as a replacement
for the initially contemplated action. Such a view implicitly defines a
notion of ``liberal efficiency'' that is analogous to the familiar concept
of Pareto efficiency.

On this latter normative view about government action, there is a broad
scope for welfare analysis but the familiar Paretian analysis is
foundationally inadequate. There are several issues in public finance, with
respect to which such a view seems to be widely held. One of these issues is
whether redistributive policy ought to be implemented by cash transfers or
by transfers in kind (e.g., provision of public housing). One of the common
arguments in favor of cash transfers is that it is demeaning to the
recipients of transfers not to be granted autonomy over the use of the
resources that society is willing to transfer.\footnote{Conversely, one of the
arguments sometimes made in favor of transfers in kind is that some classes
of recipients (e.g., addicts) are incapable of exercising such autonomy.}
This argument makes an implicit appeal to the liberal principle. It is clear
that some of the proponents of the argument regard it as being conceptually
distinct from the argument that transfers in kind are inefficient, although
those proponents (especially when they are economists) typically produce the
latter argument as a corollary. It might be thought that there could be
transfer programs with respect to which the argument about recipients'
autonomy could be raised although cash transfers would not be Pareto
superior to transfers in kind. The results of this paper indicate the
special features that such a program would have to possess. In cases where
such features are absent, an efficient transfer never violates the
preferences of recipients about how the resources dedicated to them should
be used.

It would be widely agreed that the assumptions about interests and
preferences that have been discussed here are reasonable ones to make in the
context of an evaluation of transfers to competent adults. The one
assumption from which there would possibly be substantial dissent is the
additive separability of preferences in interests. As was shown in the
preceding section, this assumption will not be satisfied if people's
preferences reflect maximin considerations regarding the satisfaction of
interests. However, the example studied in that section makes it plausible
that the results proved here are robust to this specific kind of failure of
the separability assumption. This is a matter for further research. If these
results are indeed robust, then it will be fair to conclude that the welfare
evaluation of transfers to competent adults can safely focus on efficiency
questions to the exclusion of questions about liberal choice.\footnote{It might
still be argued that there is something ethically preferable about letting
recipients make their own choices, rather than making choices for them that
are consistent with their preferences. Nevertheless, this argument does not
seem as compelling as an argument that people's preferences about matters
that are their own business are actually being thwarted. For example, most
people would judge that rather small reductions of administrative cost
constitute sufficient grounds to justify centralized allocation if the
result is consistent with recipients' preferences.}

Another issue concerns the regulation of markets for services such as
education or medical care. Because these services are differentiated
products, it is conceivably efficient to limit the number of product
varieties in order to take advantage of economies of scale. Suppose, for
simplicity, that any feasible allocation involving production of more than
one product variety is Pareto dominated by some feasible allocation in which
only a single variety is produced. Thus, according to either the efficiency
criterion or the liberal criterion, only an allocation with a single variety
can be optimal. The question is, which varieties may be produced in optimal
allocations? Parallel to the case of transfers just discussed, the answer to
this question is the same for both criteria if the assumptions studied in
this paper hold. Also parallel to that case, the appropriateness of
separability and cardinality assumptions is open to question on account of
features of the situation to which the results of this paper plausibly are
robust. However, in the case of differentiated-product allocation, there are
also intrinsic difficulties about the publicity of intrapersonal cardinal
interest comparisons. That is, the failure of the cardinality assumption to
hold may not be simply a by-product of a violation of the separability
assumption.

To understand the nature of these possible intrinsic difficulties, consider
some of the troublesome and divisive questions about resource allocation in
education and medicine. How much emphasis should schools give to the
development of students' skills in reading and arithmetic, versus the
development of critical thinking and appreciation of the arts? How much of
medical research funding should be spent on finding cures for rare diseases
that strike people early in life, versus common diseases of older people?
Debates about these questions have to do partly with distributional issues,
but they also reflect pronounced differences in the participants' views
about the relative importance of disparate elements of the good life. That
is, they reflect the fact that participants are (at least implicitly)
expressing cardinal judgments about the components of persons' interests,
and that they disagree about those judgments although their ordinal
judgments may be unanimous. It is not at all clear whether the results of
this paper can be extended to cover situations that possess this
complication.

These examples illuminate the philosophical significance of the technical
results that have been derived here. These results constitute a limited
defense of the study of the unanimity relation in welfare analysis to
address concerns that may fundamentally be about respect for persons'
objective interests. This defense acknowledges the legitimacy of these
concerns, but argues that they actually are addressed, even though
implicitly, by the technically simpler efficiency analysis. Following Mill,
a theoretical concept of a person's non-preferential interest has been
introduced in order to relate the specification of rights and liberties to
the specification of preferences. While there may be many possible ways to
give a substantive definition of this concept, only a few qualitative
assumptions about the concept are needed in order to make the defense. For
the most part, these assumptions seem to be appropriate for the discussion
of distributional issues and of other issues with which applied welfare
analysis typically deals. A formal example has shown that the assumption
that is likely to be regarded as the most restrictive one, separability, may
be stronger than is actually needed. The two economic examples that have
just been discussed show that, even if this conjecture is true, the limited
defense of the unanimity criterion falls short of showing that explicit
consideration of the liberal criterion would be completely dispensable for
applied welfare analysis.

\section*{References}

\hangindent=36pt Bergstrom, T., (1971), ``Interrelated Consumer Preferences
and Voluntary Exchange,'' {\sl Papers in Quantitative Economics}, vol. I, A.
Zarley, ed., University Press of Kansas: Lawrence, KS.

\hangindent=36pt Bergstrom, T., (1988), ``Systems of Benevolent Utility
Interdependence,'' University of Michigan, working paper.

\hangindent=36pt Bergstrom, T., (1989), ``Puzzles:  Love and Spaghetti, the
Opportunity Cost of Virtue,'' {\sl Journal of Economic Perspectives,} {\bf
3(2)}, 165-173.

\hangindent=36pt Bourbaki, N., (1965), {\sl Topologie Generale}, Hermann:
Paris, France.

\hangindent=36pt Buchanan, J., and G. Tullock, (1962), {\sl The Calculus of
Consent,} University of Michigan Press:  Ann Arbor, MI.

\hangindent36pt Debreu, G., (1959), {\sl Theory of Value,} Yale University
Press:  New Haven, CT.

\hangindent=36pt Debreu, G., (1960), ``Topological Methods in Cardinal
Utility Theory,'' in {\sl Symposium on Mathematical Methods in the Social
Sciences}, Arrow and Suppes, eds., Stanford University Press:  Stanford, CA.

\hangindent=36pt Green, E., (1979), ``A Welfare Theory for Nonmalevolent
Agents,'' Princeton University, working paper.

\hangindent=36pt Green, E., (1982), ``Equilibrium and Efficiency Under Pure
Entitlement Systems,'' {\sl Public Choice,} {\bf 39}, 185-212. [Errata
  published in {\sl Public Choice,} {\bf 41}, (1983), 237-238.] 

\hangindent=36pt Krantz, D., R.D. Luce, P. Suppes, and A. Tversky, (1971),
{\sl Foundations of Measurement,} Academic Press: New York, NY.

\hangindent=36pt Mill, J.S., (1859), {\sl On Liberty,} Parker: London,
England.

\hangindent=36pt Pareto, V., (1909), {\sl Manual of Political Economy},
Second Edition, translated by A. Schwier and edited by A. Schwier and A.
Page, reprinted by Augustus M. Kelley: New York, NY, 1971. (The translation
is of the fifth edition (1927) which is substantially identical to the
second edition.)

\hangindent=36pt Pearce, D., (1983), ``Two Essays in the Theory of Strategic
Behaviour,''  Princeton University, Ph.D. dissertation.

\hangindent=36pt Rawls, j., (1971), {\sl A Theory of Justice,} Harvard
University Press: Cambridge, MA.

\hangindent=36pt Riley, J., (1988), {\sl Liberal Utilitarianism: Social
Choice Theory and J.S. Mill's Philosophy,} Cambridge University Press:  New
York, NY.

\hangindent=36pt Rockafellar, R.T., (1970), {\sl Convex Analysis,} Princeton
University Press: Princeton, NJ.

\hangindent=36pt Schick, F., (1980), ``Towards a Logic of Liberalism,'' {\sl
Journal of Philosophy,} {\bf 77}, 80-98.

\hangindent=36pt Sen, A.K., (1970), ``The Impossibility of a Paretian
Liberal,'' {\sl Journal of Political Economy,} {\bf 78}, 152-157.

\hangindent=36pt Wicksell, K., (1935), {\sl Lectures on Political Economy,}
Routledge and Sons:  London, England.

\end{document}